\documentclass[aps,pra,superscriptaddress,twocolumn]{revtex4}
\usepackage{graphicx}
\usepackage{amssymb}
\usepackage{subfigure}
\usepackage{amsmath}
\usepackage{amsfonts}
\usepackage{ulem}
\usepackage{bm}
\usepackage{color}
\usepackage[toc,page]{appendix}
\newcommand{\ket}[1]{\ensuremath{\left|{#1}\right\rangle}}
\newcommand{\bra}[1]{\ensuremath{\left\langle{#1}\right |}}

\begin{document}

\title{Experimental demonstration of phase estimation advantage in presence of depolarizing noise by using coherent measurements. }

\author{R. S. Piera}
\affiliation{Instituto de F\'{\i}sica, Universidade Federal do Rio de Janeiro, Caixa Postal 68528, Rio de Janeiro, RJ 21941-972, Brazil}
\author{S. P. Walborn}
\affiliation{Instituto de F\'{\i}sica, Universidade Federal do Rio de Janeiro, Caixa Postal 68528, Rio de Janeiro, RJ 21941-972, Brazil}
\affiliation{Departamento de F\'{\i}sica, Universidad de Concepci\'on, 160-C Concepci\'on, Chile}
\affiliation{Millennium Institute for Research in Optics, Universidad de Concepci\'on, 160-C Concepci\'on, Chile}
\author{G. H. Aguilar}
\affiliation{Instituto de F\'{\i}sica, Universidade Federal do Rio de Janeiro, Caixa Postal 68528, Rio de Janeiro, RJ 21941-972, Brazil}

\begin{abstract}
We report an experimental investigation of the role of measurement in quantum metrology when the states of the probes are mixed. In particular, we investigated optimized local measurements and general global projective measurements, involving entangling operations, on noisy Werner states of polarization entangled photons. We demonstrate experimentally that global measurement presents an advantage in parameter estimation with respect to the optimized local strategy.  Moreover, the global strategy provides unambiguous information about the parameter of interest even when the amount of noise is not well characterized.  This shows that the coherence in quantum operations, such as the Bell-state projection device used in our protocol, can be used to further boost the quantum advantage in metrology and play a fundamental role in the design of future  quantum measurement devices.
\end{abstract}

\maketitle

\section{Introduction}

Quantum metrology is essential to improve the estimation of unknown physical parameters, and consequently, crucial for the development of new technologies and  fundamental advances \cite{Giovannetti11, Review1,LIGO1,Review2}. By using  quantum resources, such as entangled states and measurement involving entangling gates, the precision of parameter estimation can surpass the limits achieved via classical techniques \cite{Giovannetti04,Polino20}.

\begin{figure}
\includegraphics[width=8.5cm]{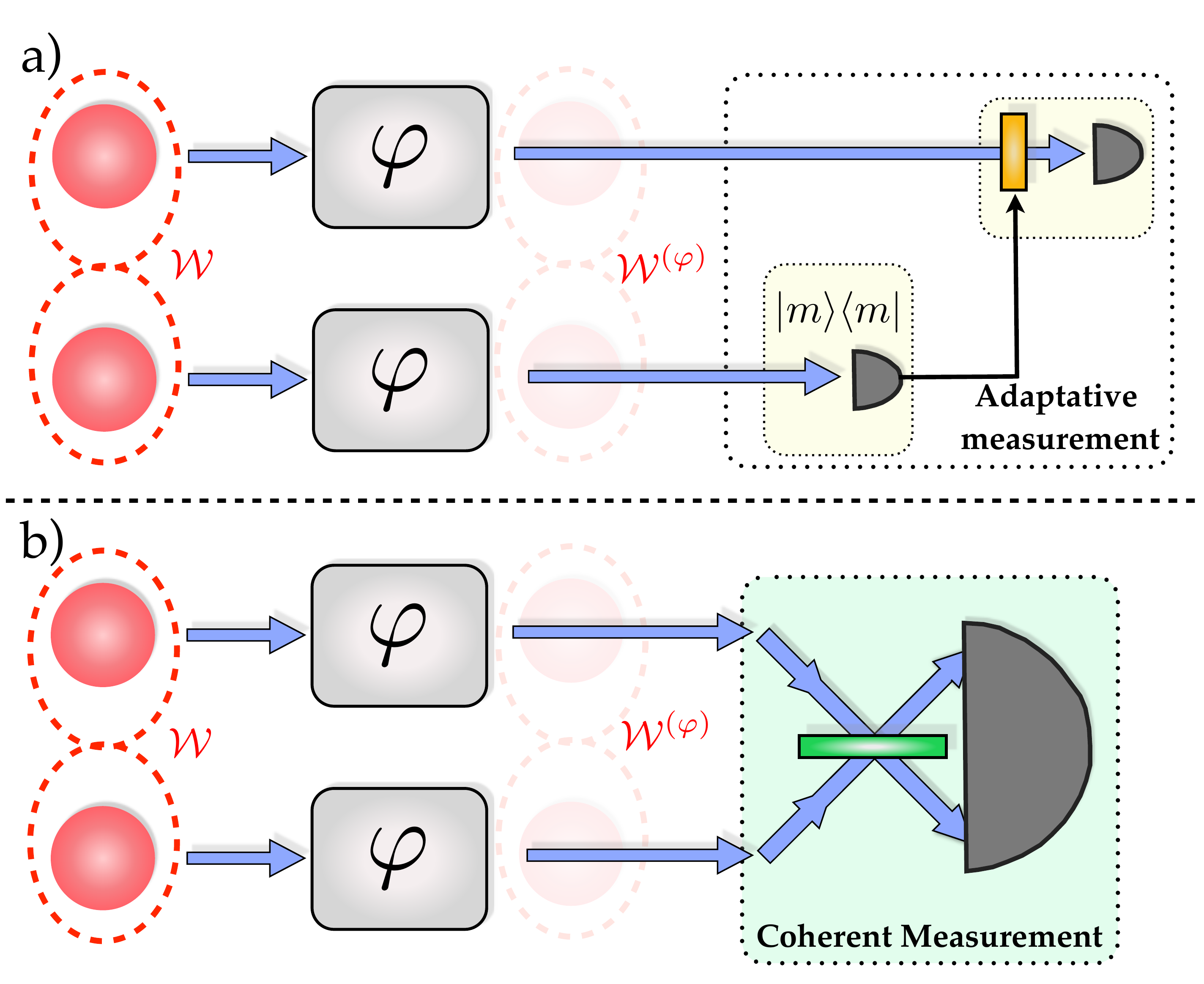}
\caption{Different measurement strategies for parameter estimation using probes in initial states $\mathcal{W}$ that may be entangled or not. a) Local adaptive measurement is performed. The state of the upper system is measured depending on the result obtained for the lower system, such that the FI is maximized.   b)  Coherent measurements involving entangling operations, such as a Bell state measurement are performed. This scheme can usually saturate the QFI when the state $\mathcal{W}$ is mixed.}
\label{fig:estimation}
\end{figure}

 A general protocol for the estimation of a parameter $\varphi$ is depicted in Fig. \ref{fig:estimation}.  A set of quantum systems--or ``quantum meters"-- are prepared in a state $\mathcal{W}$ and submitted to a process that alters $\mathcal{W}$,  imprinting the information about the parameter we want to estimate. The resulting state of the meters  $\mathcal{W}^{(\varphi)}$ is measured and the information about the parameter is extracted. The precision in the estimation of $\varphi$ depends on the choice of the meters state $\mathcal{W}$ and on the measurements strategy. Classical theory of parameter estimation predicts that the mean square error $\Delta^2 \varphi $  is lower-bounded by the classical Cram\'er-Rao  limit given by $\Delta^2\varphi\ \geq 1/NF(\varphi)$ \cite{Cramer}, where  $N$ is the number of repetitions of the experiment and $F(\varphi)$ is the Fisher information \cite{Fisher}. The Fisher information (FI) gives the amount of information about the parameter $\varphi$ that one can extract given a physical scenario. The larger the FI, the more precision one expects.  
 
 The FI is usually obtained from a parameter-dependent probability distribution $p_{\varphi}(x)$ that, in a quantum mechanical scenario, relies on the input probe state and the measurement  strategy. In this framework, to maximize the FI, one must optimize over all the measurement strategies \cite{Giovannetti11,Giovannetti06}, obtaining the quantum Fisher information $\mathcal{F}$ and the ultimate limit on the precision, given by the quantum Cram\'er-Rao  limit  $(\delta\varphi)^2 \geq 1/N\mathcal{F}$ \cite{Braunstain94}. Thus, studying the role of the measurement  for the quantum-precision enhancement is essential to obtain considerable advantages with respect to classical strategies \cite{Giovannetti11,Giovannetti06}. 
 \par
The authors of Ref. \cite{Giovannetti06} studied several protocols that differ in the amount of quantum resources present in the state $\mathcal{W}$ and in the way that the measurements are performed.  The  precision limit of the protocols involving only separable states, either be pure or mixed, scales as  $1/\sqrt{N}$ independently of the measurement scheme chosen.   When the state $\mathcal{W}$ is entangled, the limit scales as $1/N$ and doesn't depend on the measurement strategy when the states are pure.
 


 However,  pure states are impossible to produce  with real world systems.   Quantum metrology in realistic scenarios, such as in the presence of loss, noise, dissipation, or general interactions between the meter systems and the environment has been widely studied \cite{Escher11, Demkowicz09, Datta11,Lloyd08,Aguilar19}. Different strategies have been designed to fight against the detrimental effects of the environment and to protect the fragile  quantum enhancement.  For instance, some of us showed recently how the inherent phase instabilities in single-photon interferometers, used for the estimation of the tilt angle of an object, can be overcome by utilizing two photons interference \cite{Aguilar20}. The performance of local and coherent measurement in noisy quantum metrology scenarios have been investigated in Ref. \cite{micadei15}.  Error correction codes  have  been  proposed to improve the precision and to counter the effects of the noise \cite{Kessler14, Durr14, Demkowicz17, Zhou18}. In this context, introducing noiseless ancillary systems, entangled with the meters, is shown to be beneficial in the estimation when compared with single systems  \cite{Huang16, Huang18}. The measurements performed in these scenarios are  joint projections involving entangling gates. However, these kind of projections are usually hard to perform in multipartite scenarios and sometimes are not even realizable.  To circumvent such problems, several questions may be asked. For example, is it possible to perform simpler measurements and reach results close to the metrological limits? What is the best precision that can be achieved by implementing local measurements when the meters are in a mixed state? How can we improve this using entangling operations or measurements?  
 
 In this work, we answer these questions for the particular example of meters in a Werner state.  We study the precision in the estimation in the presence of noise  for separable measurements and measurements that include entangling operations. We follow the approach in Ref. \cite{micadei15}, where mixed states are considered and precision for local and global measurements are investigated. We experimentally study the estimation of a parameter creating pairs of entangled photons in a mixed state using spontaneous parametric down conversion, and implementing local projective measurements as well as a global Bell state projection using linear optics devices.  We observe an improvement in precision for the global strategy when compared to the local one.  Moreover, we note that the global strategy is capable of providing unambiguous information about the parameter even when the amount of noise is not well-characterized.

 The paper is organized as follows. In section \ref{sec:estimation} we introduce the quantum metrology framework, defining the FI and QFI. In section \ref{sec:comparition}, we discuss what are the ultimate limits when local and entangling measurement have been performed in the case that the state of the probes is a Werner state. We also  calculate the FI when the state of the photons is projected onto a Bell state and find the values of $\varphi$ in which this FI saturates the QFI. In section \ref{sec:experiment}, we describe our experimental setup and show the experimentally  reconstructed Werner states. In section \ref{sec:results}, we report and discuss the experimental results for global and local measurements. We first measure the probabilities as a function of the parameter  $\varphi$ and then calculate the FI for both cases, global and local strategies.


\section{Quantum and classical parameter estimation}
\label{sec:estimation}

 The Fisher information is one of the most important mathematical tools in the framework of quantum metrology. It is defined as follows \cite{Fisher}:
\begin{equation}
\mathcal{F}=\sum_x p_{\varphi}(x)[\partial_{\varphi} \text{ln} \left(p_{\varphi}(x) \right)  ]^2,
\label{eq:classicalfisher}
\end{equation}
where $p_{\varphi}(x)$ is the parameter-dependent probability that governs the distribution of $M$ values of $x$ obtained from some set of measurements. The FI measures the amount of information that a certain observable can extract from a physical system about a parameter $\varphi$.

  In the quantum scenario, one collects information about the parameter by performing quantum measurements, described generally by a positive operator valued measurement (POVM) $\{\mathrm{M}_x \}$, with $\int_x \text{M}_x=\mathbb{I}$,  on the state $\mathcal{W}^{(\varphi)}$.  The theoretical probabilities are related to these measurement operators via 
 \begin{equation}
 p_{\varphi}(x)=\text{Tr}\left(M_x \mathcal{W}^{(\varphi)} \right), 
\label{eq:probabilities}
\end{equation} 
and can be estimated experimentally from the frequency of the different measurement outcomes. The quantum Fisher information $\mathcal{F}$, is obtained after a hard optimization over all possible POVMs,  finding the measurement scheme that maximize the precision. This illustrates how the role of the  measurement  is crucial to improve the precision in parameter estimation. Moreover, there is always a measurement set that maximizes Eq.  (\ref{eq:classicalfisher}),  giving rise to  $\mathcal{F}$ \cite{Braunstain94}. In the special case when the parameter is imprinted in the state via a unitary operation $U$, it can be expressed as
\begin{equation}
\mathcal{F}_{Q}=2\sum_{ij}\frac{(\lambda_i-\lambda_j)^2}{\lambda_i+\lambda_j}|\langle{\psi_i}|H|\psi_j\rangle |^2
\label{eq:Quantum_fisher}
\end{equation} 
where $\{ \lambda_i \}$ and $\{ \ket{\psi_i} \}$ are the eigenvalues and eigenvectors of the state $\mathcal{W}$, respectively, $H$ is the Hamiltonian generator of the 
transformation $U=\exp(iH\varphi)$ that imprints the information of the parameter into the state as $\mathcal{W}^{(\varphi)}=U\mathcal{W}U^{\dagger}$.

\section{Entangling measurement versus best local measurement}
\label{sec:comparition}

Let us now analyze the role of measurement for a particular example where two probes ($N=2$) are in Werner state given by:
\begin{equation}
\mathcal{W}=\left( 1- \eta \right) \frac{\mathbb{I}}{4}+ \eta\ket{\mathcal{B}_{00}}\bra{\mathcal{B}_{00}},
\label{eq:werner_state}
\end{equation} 
where  $\ket{\mathcal{B}_{00}}$ is one of the four maximally entangled Bell states, defined as 
\begin{equation}
\ket{\mathcal{B}_{kj}}=1/\sqrt{2}\left(\ket{0,j}+(-1)^{k}\ket{1,1\oplus j}\right),
\label{eq:Bellstates}
\end{equation}with integers $k,j$ taking values 0 or 1. 
The noise parameter $\eta$ varies between 0 and 1. This state $\mathcal{W}$ is obtained when a pair of qubits, initialized in the state $\ket{\mathcal{B}_{00}}$, is submitted to a depolarizing channel. We note that any mixed state can be transformed into a state of the form \eqref{eq:werner_state} using random bilateral rotations \cite{bennett96}. For this state, it is well known that the best (worst) parameter estimation is obtained for $\eta=1$ ($\eta=0$),  when quantum correlations are maximum (minimum). We consider that the Hamiltonian in Eq. (\ref{eq:Quantum_fisher}) is a sum of local Hamiltonians given by $H =\oplus_i H_i $, where  $H_i=\ket{1}\bra{1}$. Thus,  the phase $\varphi$ is imprinted only when probes are in state $\ket{1}$, leaving the probes in state $\ket{0}$ unchanged. The resulting state after the interaction can be written as:  
\begin{eqnarray}
\mathcal{W}^{(\varphi)}&=&\left( 1- \eta \right) \frac{\mathbb{I}}{4} \nonumber \\
&+& \frac{\eta}{2}\left(\ket{00}+e^{i2\varphi}\ket{11}\right)\left(\bra{00}+e^{-i2\varphi}\bra{11}\right).
\label{eq:werner_state2}
\end{eqnarray} 
For the general case, which includes coherent measurements, $\mathcal{F}$ can be calculated using Eq. (\ref{eq:Quantum_fisher}), giving \cite{micadei15}
\begin{equation}
\mathcal{F}^{co}(\mathcal{W})=\frac{8\eta^2}{1+\eta},
\label{eq:Fisher_global_theory}
\end{equation} 
which corresponds to the ultimate precision in the parameter estimation. The superscript \textit{co} means coherent measurements that could involve entangling operations. 
\par
We turn now to study the precision limits when projective measurements onto the maximally entangled Bell basis are performed. Using Eqs.  \eqref{eq:Bellstates} and \eqref{eq:werner_state2}, the associated probabilities are
\begin{eqnarray}
P_{\mathcal{B}_{00}}&=&\frac{1+\eta}{4}+\frac{\eta\cos(2\varphi)}{2},\nonumber \\
P_{\mathcal{B}_{10}}&=&\frac{1+\eta}{4}-\frac{\eta \cos(2\varphi)}{2},\label{eq:prob_fitting2}\\
P_{\mathcal{B}_{01}}&=&P_{\mathcal{B}_{11}}=\frac{1-\eta}{4}, \nonumber
\end{eqnarray}
where $P_{\mathcal{B}_{kj}}$ is the probability when a projective measurement onto $\ket{\mathcal{B}_{kj}}$ have been performed. One can see that probabilities of having photons in the state $\ket{\mathcal{B}_{11}}$ and $\ket{\mathcal{B}_{01}}$ are independent of the parameter $\varphi$. Thus, these states do not contribute to the estimation of the parameter $\varphi$. 
Using the probabilities in Eq. (\ref{eq:probabilities}) and the formula for the FI (\ref{eq:classicalfisher}), we obtain 
\begin{equation}
F^{Bell}(\mathcal{W})=\frac{8\eta^2 (1+\eta)\sin^2\left( 2 \varphi \right)}{(1+\eta)^2-4\eta^2 \cos\left( 2 \varphi \right)},
\label{eq:Fisher_Bell}
\end{equation} 
 which saturates the quantum Fisher information $\mathcal{F}^{en}$ when the reference value of the parameter is $\varphi=\pi/4$. Given that derivative of $F^{Bell}$ at  $\varphi=\pi/4$ is zero,  there is a small interval of values of $\varphi$, close to $\pi/4$,  where $F^{Bell}$ is very close to the QFI in Eq. (\ref{eq:Fisher_global_theory}).  

\begin{figure}
\includegraphics[width=8.7cm]{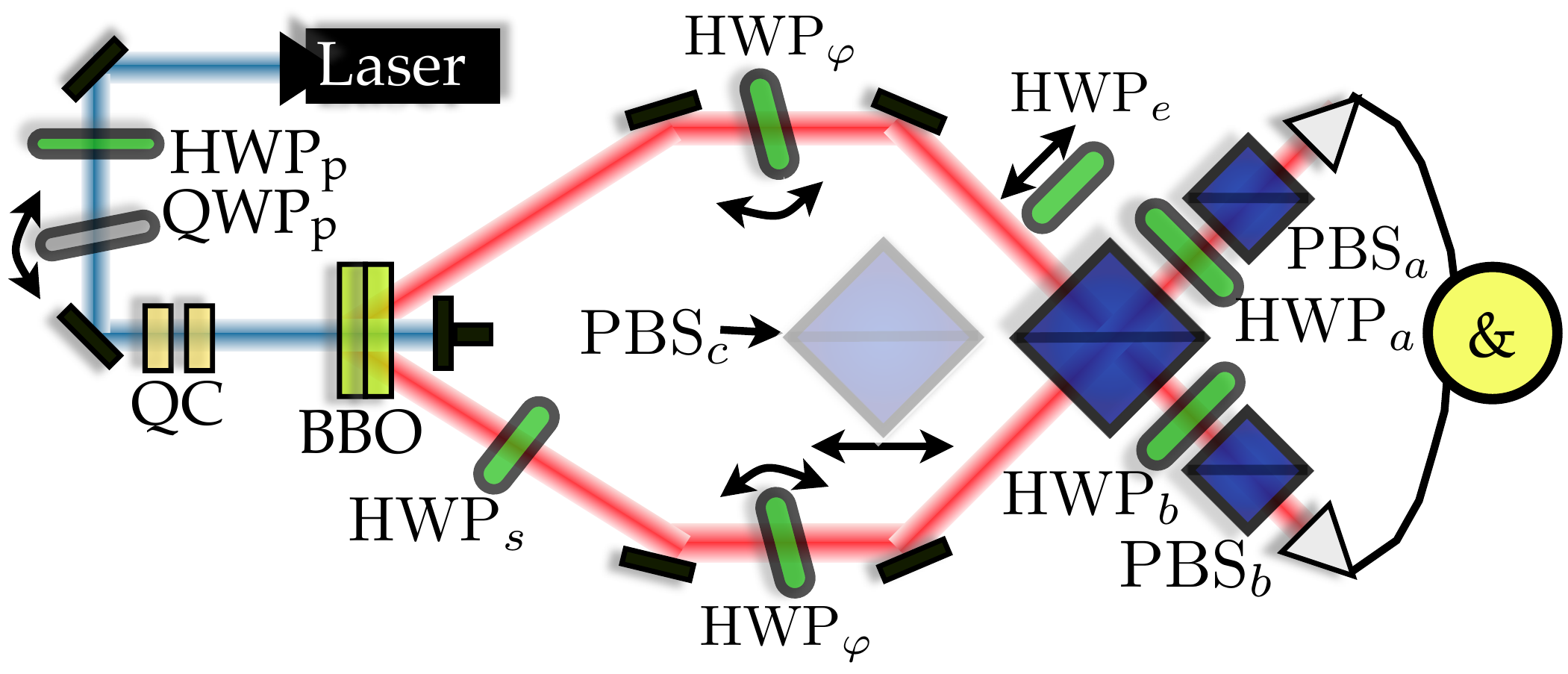}
\caption{a)Experimental setup: Photon pairs emerge from the BBO crystal in a state that can be entangled or not depending on the configuration of HWP$_p$. To obtain Werner states, QWP$_p$ and HWP$_s$ are varied during the measurement time. The  parameter to be estimated is the phase $\varphi$ between the horizontal and vertical polarizations, which is altered by tilting the plates HWP$_{\varphi}$. We can toggle between global and local projective measurements by inserting or removing PBS$_c$. Projection onto different local polarization states is realized using  PBS$_i$ and HWP$_i$, with $i$ equal to $a$ or $b$.   Photons are coupled into single-mode fibers and detected by single-photon detectors.  }
\label{fig:setup}
\end{figure}

To compare the global strategy to the best local one, let us consider a different tactic in which we have access only to local measurements. The question is: ``what is the best measurement that can be performed such that a minimum uncertainty  $\varphi$ is obtained?". This question was addressed in Ref. \cite{micadei15}. The authors assumed  that the measurements on both probes can be performed adaptively, such that the second measurement can be chosen depending on the result of the first one, thus maximizing the precision of the parameter estimation.  By projecting one of the probes onto an arbitrary qubit state $\bra{m}=m_0\bra{0}+m_1\bra{1}$, the conditional state of the other probe is
\begin{eqnarray}
\mathcal{W}_{c}^{(\varphi)}&=&\left( 1- \eta \right) \frac{\mathbb{I}}{2}\\
&+& \frac{\eta}{2}\left(m_0\ket{0}+e^{i2\varphi}m_1 \ket{1}\right)\left(m_0^*\bra{0}+e^{-i2\varphi}m_1^*\bra{1}\right).\nonumber
\label{eq:werner_state_single}
\end{eqnarray} 
Notice that there is no information of the phase $\varphi$ coming from the first measurement. By calculating the quantum Fisher information using Eq. (\ref{eq:Quantum_fisher}) of the single system that was not yet measured, we obtain \cite{micadei15}:
\begin{equation}
\mathcal{F}^{adp}\left(\mathcal{W}_c\right)=16|m_0 m_1|^2\eta^2,
\label{eq:Fisher_local_theory}
\end{equation} 
where $\mathcal{F}^{adp}$ is the quantum Fisher information calculated for the remaining state of Eq. (\ref{eq:werner_state_single}), which is maximized when $|m_0|=|m_1|=1/\sqrt{2}$. Note that this QFI corresponds to the single system, being in general smaller than the QFI deduced in (\ref{eq:Fisher_global_theory}). The maximum value of $\mathcal{F}^{adp}$ is obtained by projections onto the equator of the Bloch sphere. 
For instance, for projection onto the four separable states $\ket{\pm \pm}$, where $\ket{\pm}=1/\sqrt(\ket{0}\pm\ket{1})$ are the diagonal states, the associated probabilities can be written as
\begin{eqnarray}
P_{++}&=&\frac{1+\eta \cos(2\varphi)}{4}=P_{--}, \nonumber\\
P_{+-}&=&\frac{ 1 -\eta \cos(2\varphi)}{4}=P_{-+}, 
\label{eq:prob_fitting1}
\end{eqnarray}
where for example $P_{++}$ is the probability of projecting the state of the photons onto $\ket{++}$.  The Fisher information for  measurements in this basis  is 
\begin{equation}
F^{\pm} = \frac{4 \eta^2 \sin^2(2 \varphi)}{1-\eta^2\cos^2(2\varphi)}.
\label{eq:Fdiag}
\end{equation}
  $F^{\pm}$ saturates the QFI \eqref{eq:Fisher_local_theory} for the local adaptive strategy when $\varphi$ is near $\pi/4$. We note that in this scheme the measurements are not adaptive in the sense that one measurements depend upon the results of the other.  Rather, it is simply necessary to combine the measurement results (one-way classical communication) in order to determine the joint probabilities \eqref{eq:prob_fitting1}. 
\par
Let us point out  an advantage to the Bell-state measurement, in relation to the local adaptive strategy.  For the local measurements, we can see from the probabilities \eqref{eq:prob_fitting1} as well as the FI \eqref{eq:Fdiag} that the available information about $\varphi$ is limited by our knowledge of $\eta$.  That is, if $\eta$ is unknown, or varying in time, it can be difficult or even impossible to obtain a reliable estimate for $\varphi$.  We can see that an experimentalist monitoring only the probabilities \eqref{eq:prob_fitting1} cannot reliably distinguish between a change in $\eta$ and a change in $\varphi$ unless some a priori knowledge of $\eta$ is available.  This has been vastly discussed in multi-parameter estimation scenarios \cite{Parniak18,Vidrighin14}. We note that a similar situation can also occur for almost any interferometric measurement \cite{Aguilar20, Walborn20}.  For the Bell state measurement, on the other hand, we can see from probabilities \eqref{eq:prob_fitting2} that two results return information on $\eta$ alone, while the other two depend on both $\eta$ and $\varphi$.  Thus, from the single set of measurement results one can first determine an estimate for $\eta$, and then use this to obtain a reliable estimate of $\varphi$.  We will employ this advantage in section \ref{sec:results}.
\par
In summary, following Ref. \cite{micadei15}, we have found measurement strategies that can saturate the QFI for both entangling measurements as well as optimized local (adaptive) measurements.  In the following sections, we implement these strategies for photons using linear optical devices.
\par

\begin{figure}
\centering
\includegraphics[width=8cm]{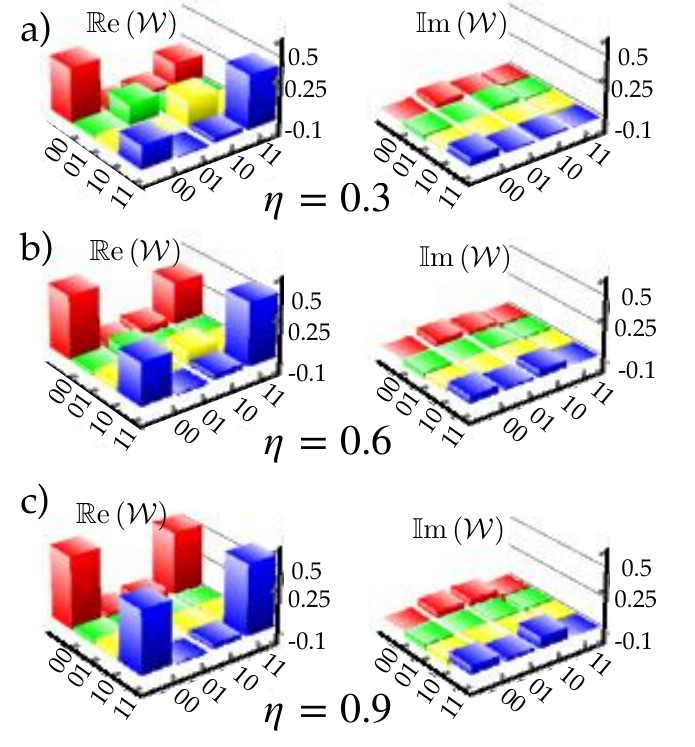}
\caption{ Real (left panels) and imaginary (right panels) parts of the reconstructed density matrices of the Werner states for different values of $\eta$. }
\label{fig:matrices}
\end{figure}

\section{Experiment}
\label{sec:experiment}

Both parameter estimation strategies described above were investigated experimentally using   entangled photons produced from spontaneous parametric down conversion (SPDC). The experimental setup is shown in Fig. \ref{fig:setup}.  A continuos-wave laser, centered at 405 nm, pumps two cross-axis beta-barium-borate (BBO) crystals, producing pairs of photons at 810 nm from type-I phasematched SPDC \cite{Kwiat99}. To obtain entangled states in the polarization degree of freedom, two birefringent quartz crystals (QC) were located in the path of the laser to compensate longitudinal polarization walk-off in the crystals \cite{Kwiat09}. HWP$_{\text{p}}$ is set to produce a state $\ket{\Phi^{\theta}}=\frac{1}{\sqrt{2}}\left[ \ket{00}+e^{i\theta}\ket{11} \right]$, where  $\ket{0}$ and $\ket{1}$ stand for the horizontal and vertical polarization, respectively.   The phase $\theta$ is controlled by tilting the angle of the QWP$_\text{p}$. With this, the four different Bell states \eqref{eq:Bellstates}  can be obtained with different configurations of  QWP$_p$ and HWP$_s$ \cite{Kwiat99}.   This allows us to generate the Werner state \eqref{eq:werner_state},  since it can be written as a convex sum of Bell states as 
\begin{eqnarray}
\mathcal{W}&=&\frac{\left( 1- \eta \right)}{4}\left(\ket{\mathcal{B}_{10}}\bra{\mathcal{B}_{10}}+\ket{\mathcal{B}_{01}}\bra{\mathcal{B}_{01}}+\ket{\mathcal{B}_{11}}\bra{\mathcal{B}_{11}} \right) \nonumber\\&+& \frac{\left( 1+3 \eta \right)}{4}\ket{\mathcal{B}_{00}}\bra{\mathcal{B}_{00}}.
\end{eqnarray}
To generate this state experimentally, we change configurations of  QWP$_p$ and HWP$_s$ during measurement acquisition time. For instance, to obtain $\mathcal{W}$ with $\eta=0.5$ we set QWP$_p$ and HWP$_s$ to produce $\ket{\mathcal{B}_{00}}  $ during 5/8 of the total acquisition time, while the remaining 3/8 of the time is equally divided among the other three Bell states, and the wave plates set accordingly.  With PBS$_c$ removed from the setup, we can use the sets of wave-plates HWP$_i$, QWP$_i$ (not included in the figure) and PBS$_i$, where $i$ can be either $a$ or $b$, to perform local projective polarization measurements on each photon \cite{james01}. Photons are detected using $\sim$60$\%$ efficiency single-photon avalanche diodes (APD) and coincidence counts are registered with $\sim 3\%$ overall detection efficiency, including collection efficiency into single mode fibers. Fig. \ref{fig:matrices} shows the tomographically reconstructed density matrices for different values of $\eta$. The fidelities $F=\sqrt{\sqrt{\mathcal{W}} \rho \sqrt{\mathcal{W}}}$   of the produced states $\rho$ with respect to the states in Eq.(\ref{eq:werner_state}) are above $93\%$ for $\varphi=0$.

To estimate the parameter using the global strategy, we need to implement a Bell state projection using linear optics devices \cite{mattle96}. This is done by placing PBS$_c$ in the path of the photons and adjusting the path length of both photons to be equal, obtaining Hong-Ou-Mandel (HOM) interference \cite{HOM}.  The interference  visibility was $V=0.96(2)$. When the outputs of PBS$_c$ are projected onto $\ket{\pm \pm}$, a global projection onto Bell states $\ket{\mathcal{B}_{00}}$ or $\ket{\mathcal{B}_{10}}$  is realized when the signs are the same or different, respectively.    In order to project onto the other two Bell states, HWP$_e$ is introduced in the setup transforming $\ket{0}\rightarrow\ket{1}$, and thus mapping  $\ket{\mathcal{B}_{k0}}$ onto $\ket{\mathcal{B}_{k1}}$ \cite{Aguilar12}.  

\begin{figure}
\includegraphics[width=8.7cm]{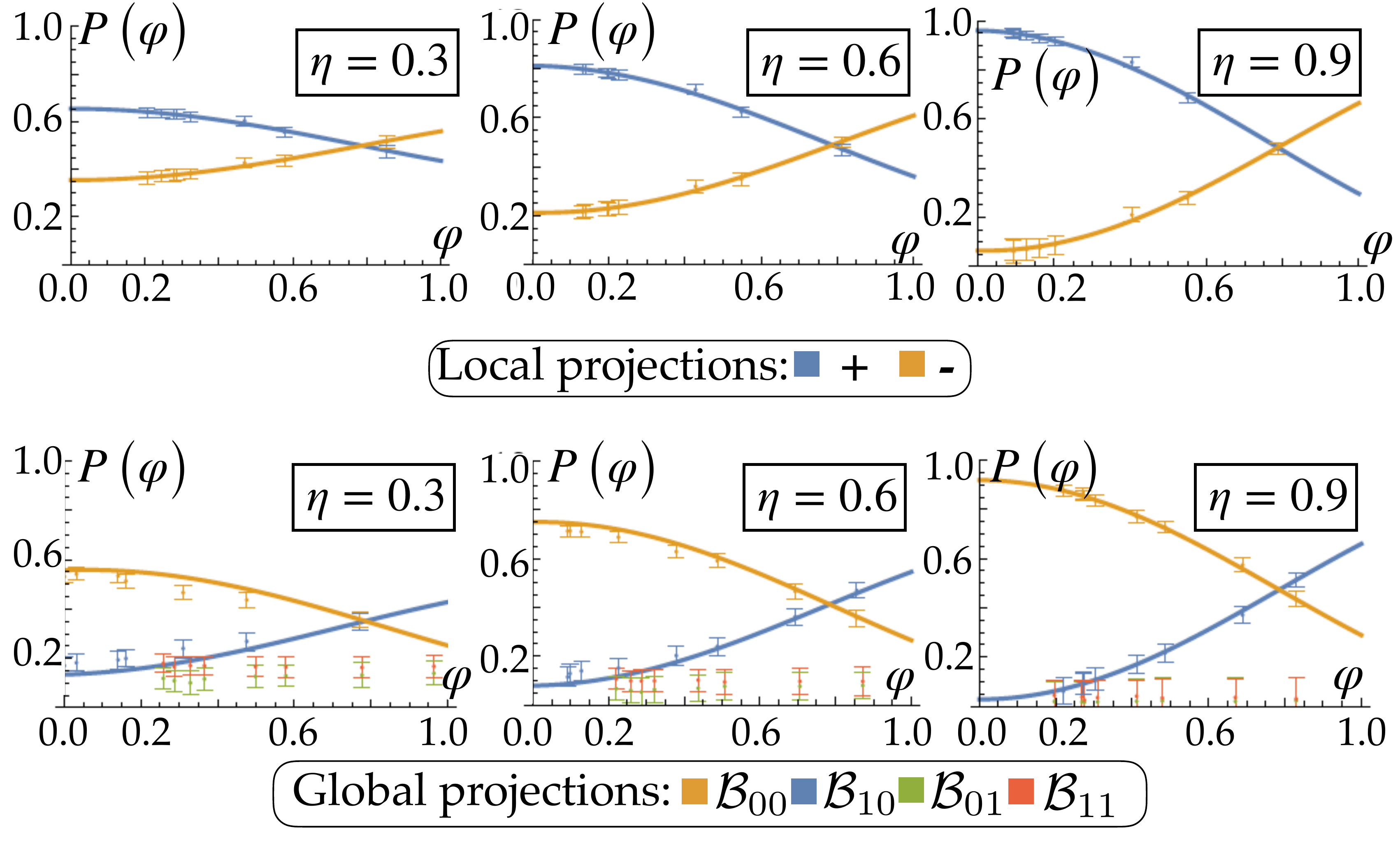}
\caption{a)Experimental probabilities as function of $\varphi$ for different values of $\eta$. The upper three plots corresponds to the probabilities $P_+=P_{++}+P_{--}$ and $P_-=P_{+-}+P_{-+}$ for the local strategy, while the lower plots are probabilities for the  global strategy. The lines are fittings obtained from the Eq.  Eq. (\ref{eq:prob_fitting1}) and (\ref{eq:prob_fitting2}) multiplying the cosine terms by a factor V. See main text for more detail.}
\label{fig:probabilities}
\end{figure}

The parameter  $\varphi$ to be estimated is a phase  between the horizontal and vertical polarization introduced by two identical HWP$_{\varphi}$.  Different values of $\varphi$ are obtained by changing the tilt angle of these plates. The parameter estimation using the adaptive local strategy was realized using the set PBS$_i$ and HWP$_i$ utilized before for state tomography. We set the  angles of HWP$_i$ to performs projection of the photons onto states in the equator of the Bloch sphere. 

\section{Results}
\label{sec:results}

\begin{figure}
\includegraphics[width=8.7cm]{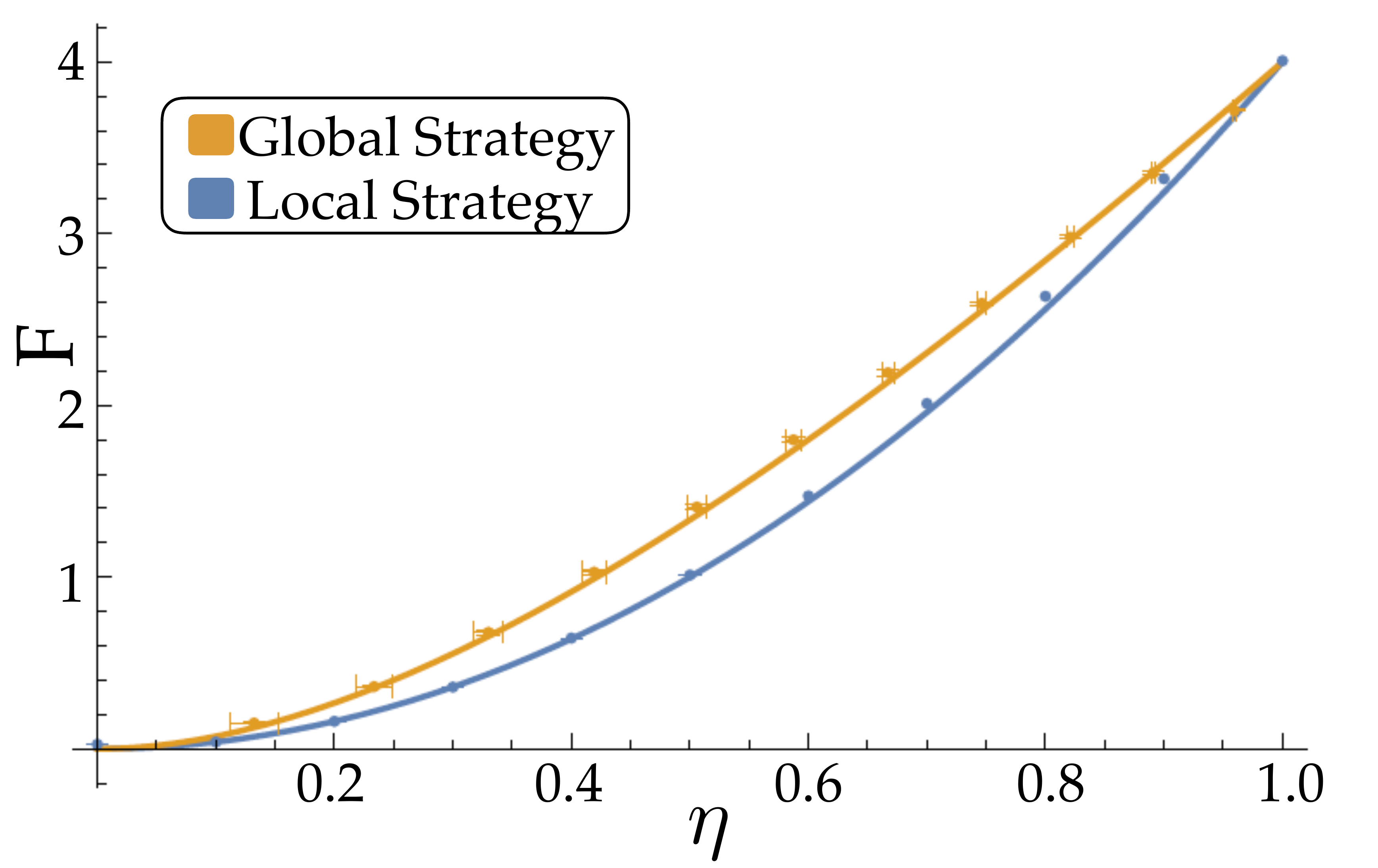}
\caption{a) Fisher information for local and global strategies as a function of $\eta$ for $\varphi=\pi/4$. Orange and blue dots corresponds to experimental values of $\mathcal{F}$ for local and global, respectively. The theoretical predictions, given by equations (\ref{eq:Fisher_local_theory}) and (\ref{eq:Fisher_global_theory}), are shown in solid orange and blue lines, respectively.  Experimental data coincide with the theoretical curves within the error bars. }
\label{fig:fisherinfo}
\end{figure}

We measure the probability of finding  photon pairs in each projection,   local and global,  while $\varphi$ is varied. The results are shown in figure \ref{fig:probabilities}.  Given that for the local projection in Eq. (\ref{eq:prob_fitting1}) the probabilities $P_{++}$ and $P_{--}$ ($P_{+-}$ and $P_{-+}$ ) have the same behavior,  we plot $P_+=P_{++}+P_{--}$ and $P_-=P_{+-}+P_{-+}$. For the global projection, one can see that $P_{\mathcal{B}_{11}}$ and $P_{\mathcal{B}_{01}}$ are independent of the values of $\varphi$. A  discrepancy between  experiment and theory is observed when trying to fit the experimental data with the expressions in Eq. (\ref{eq:prob_fitting1}) and (\ref{eq:prob_fitting2}).  For example, for $\eta=1$ unit  visibility curves are expected, which is not observed  experimentally. This is mostly related with  experimental imperfections, such as the production of non-unit purity  Bell states  and  small mismatching of the photons at PBS$_c$, which generate a small distinguishability of the photons, reducing the visibility of the HOM interference required for projection onto Bell states.  To account for this experimental issue, we added the visibility parameter $V=0.96(2)$ multiplying the cosine terms in Eq. (\ref{eq:prob_fitting1}) and (\ref{eq:prob_fitting2}), which allowed for excellent agreement between theory and experiment for both strategies. We can observe that the visibilities of the oscillations in Fig. \ref{fig:probabilities} increase while $\eta$ increases, in agreement with what is expected.  In Figs. \ref{fig:probabilities} and \ref{fig:fisherinfo}, error bars were calculated by error propagation using the Poissonian coincidence count statistics. 

As discussed near the end of section \ref{sec:comparition}, there can be ambiguity concerning the value of $\eta$ for the local strategy.  This ambiguity can be removed in the global approach.  As such,  the parameter $\eta$ is estimated differently for  local and global strategies. For the former, we  determine $\eta$ by the relations of measurement time in each configuration of  QWP$_p$ and   HWP$_s$ as was done in the tomographic measurements, as discussed in section \ref{sec:experiment}. For the global case, $\eta$ is estimated from the probabilities $P_{\mathcal{B}_{01}}$ and $P_{\mathcal{B}_{11}}$ that do not depend on the parameter $\varphi$.  For both strategies, $\varphi$ was estimated using a Likelihood method, where the probabilities predicted by the model are in Eq. (\ref{eq:prob_fitting1}) and (\ref{eq:prob_fitting2}) and the frequencies are the number of coincidences obtained in each case.

Using the curve fits $p^\prime(\varphi,\eta)$ obtained from the probability curves  in Fig. \ref{fig:probabilities}, we applied  Eq. (\ref{eq:classicalfisher}) to obtain an experimental estimate of the Fisher information for both local and global strategies. The results are shown in Fig. \ref{fig:fisherinfo} as a function of $\eta$.  A small discrepancy between the theory and experiment is observed for $\eta$ close to 1, due to the fact that the HOM interference visibility never reaches unity. However, the net effect of this is smaller than the error bars.   One can observe that the experimental $F$ for the global strategy is  larger than $F$ for the local strategy for all values of $\eta$ except for $\eta=0$ and $\eta=1$, where they coincide. 
 This is because there is no local measurement  that reaches the QFI \eqref{eq:Fisher_global_theory} when the input state is a Werner state.  This result is general when the states of the probes are mixed \cite{micadei15, Modi11}. For probes in a pure states, both kind of measurements give the same precision \cite{Giovannetti06}. 
 \par
We note that the linear-optics based global measurement device used here is capable of demonstrating a quantum advantage, even though it is impossible to project onto all four Bell states using linear optics \cite{lutkenhaus99,calsamiglia01}. This is due to the fact that here it is only necessary to distinguish the three groups $\{ \mathcal{B}_{00}\}$, $\{\mathcal{B}_{10}\}$ and $\{\mathcal{B}_{01},\mathcal{B}_{11}\}$.  This can be done completely in a single measurement device using photon-number sensitive detectors \cite{mattle96}.  To avoid this requirement for the sake of investigation, we have employed the measurement strategy above, detecting two Bell states at a time.  Our scheme can be employed directly in a static configuration by using four single photon detectors and accepting a 50\% efficiency in detection of group $\{\mathcal{B}_{01},\mathcal{B}_{11}\}$ (losing events where two photons go to the same detector) which only provides information on $\eta$.  Similarly, we could detect the three groups above with 100\% efficiency (in theory) with single photon detectors by altering the spatial profile of the pump laser \cite{walborn03b}. 
\par
It is also important to notice that the depolarizing channel and the imprinting of a relative phase do not commute in general.  However, when the initial state of the meters is $\ket{\mathcal{B}_{00}}$, imprinting a relative phase and then passing through a depolarization channel gives the same result as imprinting a phase on the depolarized state. This opens the possibility of some practical applications where the information of the parameter can be encoded before, after or in the midst of depolarizing noise. For a very relevant practical example, let us consider that we want to estimate a phase between the polarization components, introduced by a birefringent material, for example, placed along some long-distance optical channel. The channel could be realized by either optical fibers or a free-space propagation. It is known that the transmission in optical fibers depolarizes the input state of the photons (time averaged), i. e. different polarization components travel with distinct group velocity (birenfringence), as well as depolarization noise produced by Raman scattering. Thus, if we prepare the photons in a superposition of horizontal and vertical polarizations, after their propagation through the fiber, they arrive at the measurement station in a state with a certain degree of mixture (time averaged). When entangled photons are being used to improve the estimation of this phase, the photons would arrive at the measurement station in a Werner state similar to those of the Eq. (\ref{eq:werner_state}).  The situation is not different when photons travel through a free-space channel. For this kind of channel, the photons have to be collected with large aperture telescope, which not only collects the light of interest but also a depolarized background of ambient light,  thus degrading the polarization state.  In these both practical scenarios, performing Bell state projection, instead of local measurements, should be both beneficial, as has been shown in Fig. \ref{fig:fisherinfo}.  Moreover, our setup allows one to characterize the amount of depolarization.

\par
Let us mention that in Ref. \cite{Huang18} precise estimation of the phase introduced by a HWP has also been studied using quantum metrology protocols. In this work, the state of the photons evolve after  passing through different noisy channels. The results show that the parameter estimation is better when a noiseless ancillas entangled  with the probe are used instead of a single system, demonstrating advantages in the entanglement-assisted protocols of quantum metrology. Our result is conceptually different. We show that a realistic global measurement can be used to surpass the best known local strategy for the case of two noisy entangled probes.

\section{Conclusions}
Quantum metrology is at the heart of the important advances in applied and fundamental science as well as benefits to industry. It is of fundamental importance to understand how to boost the precision of quantum metrology using more involved, global measurements, as well as to define the limitations of simple local measurements in real scenarios. Here we show experimentally that global measurements, such as projection onto entangled Bell states, can outperform the best local strategies when the input state is a mixed Werner state with arbitrary noise parameter $\eta$. This is important to limit the precision in cases where only local measurements are accessible. Our Bell-state analyzer operates with linear optics devices based on two-photon interference at a beam splitter.  It is capable of saturating the quantum Fisher information even though it does not perform a complete projection onto the Bell basis. In addition, we identify an advantage of our global measurement strategy in that it is capable of simultaneously providing an estimate for the amount of noise  as well as the parameter of interest.  This can be advantageous when the noise is not well-characterized, or when it is time varying.

\begin{acknowledgements}
We dedicate this paper to Daniel ``el profe" Cordoba, who sadly passed away a few month ago. Thank you Daniel for transmitting your passion for physics, you will always be in our work and our memories. We thank Roberto Serra and Ruynet de Matos Filho for valuable discussions. 
The authors would like to thank CAPES, CNPQ and the INCT-IQ for partial
financial support.  This work was realized as part of the CAPES/PROCAD program.  SPW received support from the Fondo Nacional de Desarrollo Cient\'ifico y Tecnol\'ogico  (1200266) and the Millennium Institute for Research in Optics. 
\end{acknowledgements}

\bibliographystyle{apsrev}

\end{document}